\documentstyle[epsfig,12pt,a4p,subfigure]{article}

\newcommand{\pipi}{\mbox{$\pi^{+}\pi^{-}$} }

\newcommand{\pom}{$I\hspace{-1.6mm}P$}
\begin{document}
\begin{titlepage}
\def\footnoterule{\hrule width 1.0\columnwidth}
\begin{tabbing}
put this on the right hand corner using tabbing so it looks
 and neat and in \= \kill
\> {7 August 2000}
\end{tabbing}
\bigskip
\bigskip
\begin{center}{\Large  {\bf
Resonance production in central $pp$ collisions at
the CERN Omega Spectrometer}
}\end{center}
\bigskip
\bigskip
\begin{center}{
A.\thinspace Kirk
}\end{center}
\bigskip
\bigskip
\begin{tabbing}
aba \=   \kill
\> \small
School of Physics and Astronomy, University of Birmingham, Birmingham, U.K. \\
\end{tabbing}
\begin{center}{\bf {{\bf Abstract}}}\end{center}

{
A study of resonance production in central $pp$ collisions is presented
as a function of several kinematical variables.
In particular the difference in the transverse momentum ($dP_T$) of
the exchanged particles shows that undisputed $q \bar{q}$ mesons are
suppressed at small $dP_T$ whereas glueball candidates are enhanced and
in addition,
the azimuthal angle $\phi$ gives information on the nature of the Pomeron.
}
\bigskip
\bigskip
\bigskip
\bigskip\begin{center}{{Submitted to Physics Letters}}
\end{center}
\end{titlepage}
\setcounter{page}{2}

\par
The Omega central production experiments
(WA76, WA91 and WA102) were
designed to study exclusive final states
formed in the reaction
\begin{center}
pp$\longrightarrow$p$_{f}X^{0}$p$_s$,
\end{center}
where the subscripts $f$ and $s$ refer to the fastest and slowest
particles in the laboratory frame respectively and $X^0$ represents
the central system. Such reactions are expected to
be mediated by double exchange processes
where both Pomeron and Reggeon exchange can occur.
The trigger was designed to enhance double exchange
processes with respect to single exchange and elastic processes.
Details of the trigger conditions, the data
processing and event selection
have been given in previous publications~\cite{re:expt}.
\par
The experiments have been
performed at incident beam momenta of 85, 300 and 450 GeV/c, corresponding to
centre-of-mass energies of
$\sqrt{s} = 12.7$, 23.8 and 29~GeV.
Theoretical
predictions \cite{pred} of the evolution of
the different exchange mechanisms with centre
of mass energy, $\sqrt{s}$, suggest that
\begin{center}
$\sigma$(RR) $\sim s^{-1}$,\\
$\sigma$(R\pom) $\sim s^{-0.5}$,\\
$\sigma$(\pom\pom) $\sim$ constant,
\end{center}
where RR, R\pom \thinspace \thinspace
and \pom\pom \thinspace \thinspace refer
to Reggeon-Reggeon, Reggeon-Pomeron and Pomeron-Pomeron
exchange respectively. Hence we expect Double Pomeron Exchange
(DPE) to be more significant at high energies, whereas the Reggeon-Reggeon and
Reggeon-Pomeron mechanisms will be of decreasing importance.
The decrease of the non-DPE cross section with energy can be inferred
from data
taken by the WA76 collaboration using pp interactions at $\sqrt{s}$ of 12.7 GeV
and 23.8 GeV \cite{wa76}.
The \pipi mass spectra for the two cases show that
the signal-to-background ratio for the $\rho^0$(770)
is much lower at high energy, and the WA76 collaboration report
that the ratio of the $\rho^0$(770) cross sections at 23.8 GeV and 12.7 GeV
is 0.44~$\pm$~0.07.
Since isospin 1 states such as the $\rho^0$(770) cannot be produced by DPE,
the decrease
of the $\rho^{0}(770)$ signal at high $\sqrt{s}$
is consistent with DPE becoming
relatively more important with increasing energy with respect to other
exchange processes.
\par
Due to the fact that there was no electromagnetic calorimeter in
the WA76 experiment at $\sqrt{s}$ of 12.7 GeV, the $s$ dependence of
only a small number of states can be determined.
The states for which the $s$ dependence have been measured
and whose compatibility with DPE have been determined are given
in table~\ref{summary}.
Although there was no calorimeter
in the 85~GeV/c run of the WA76 experiment, it
was able to reconstruct the
$\eta \pi^+\pi^-$ mass spectrum using the
decay $\eta$~$\rightarrow$~$\pi^+\pi^-(\pi^0)_{missing}$~\cite{phiangpap}.
In this mass spectrum the $\eta^\prime$ and $f_1(1285)$ are seen.
The cross section of the $f_1(1285)$ at 85 GeV/c has been well
measured through its all charged particle decay modes
and hence can be used to determine the
cross section of the $\eta^\prime$.
\par
After correcting for
geometrical acceptances, detector efficiencies,
losses due to selection cuts,
and all known decay modes,
the cross-sections for the
production of all the resonances
at $\sqrt s$~=~29.1~GeV
observed in the WA102 experiment are shown in table~\ref{summary}.
As can be seen,
there is evidence
that in central $pp$ collisions $s \bar{s}$ production is much weaker than
$n \bar{n}$ production.
For example,
the cross section for the production of the $f_2(1270)$, whose production
has been found to be consistent with DPE, is
more than 40 times greater than the cross section of the $f_2^\prime(1525)$.
Part of this suppression could be due to the fact that there is a
kinematic suppression of $M_{X^0}^{-2}$
and some could be explained by the fact that the centre of mass energy
dependence of the $f_2(1270)$ allows for up to a 30~\% contribution
from non-DPE processes. However, a suppression of $\approx~20$ still
needs to be explained.
Hence there could be some strong dependence on
the mass of the produced quarks in DPE.
A similar effect
is observed in electroproduction ($\gamma$\pom)
where the $\phi$ becomes more
suppressed relative to the $\rho$ at low Q$^2$ of the
photon~\cite{H1Collab}.
\par
The cross section as a function of mass has not previously been
shown for the centrally produced system.
This distribution could be determined in two ways.
Either by measuring the missing mass from the two outgoing protons
or by measuring the effective mass of the decay products of the
central system. The latter method has the disadvantage of requiring a
separate acceptance correction for each final state.
However, due to the measurement errors on the outgoing fast proton,
the second method has the
overwhelming advantage of having more than an order of magnitude
better resolution in the determination of the mass.
In order to determine the
$d\sigma/dM_{X_0}$
differential cross section,
starting with the $\pi \pi$ final state,
the acceptance corrected mass spectra
of all the final states observed in the WA102 experiment
have been added together and the resulting
differential cross section,
$d\sigma/dM_{X_0}$,
is shown in
fig.~\ref{fi:1}a). Prominent features are the $\eta$, $\omega$, $\eta^\prime$,
$f_1(1285)$, $f_1(1420)$ and a broad enhancement around 2~GeV.
\par
In addition to the cross section and $s$ dependences, there are a number
of different kinematical variables that can be determined for each resonance.
The first is the Feynman $x$ ($x_F$) distribution of the central system.
It should be noted that for resonances that decay into all charged
particle decay modes the $x_F$ distribution measured is symmetric about 0.
However, for resonances that decay into final states including $\gamma$s,
because of the acceptance of the calorimeter, only events
with $x_F$~$>$~0. are detected.
Hence for consistency, only $x_F$~$>$~0 will be shown.
All the resonances observed tend to fall into two distinct classes:
fig.~\ref{fi:2}a) those that are peaked at $x_F$~=~0 and
fig.~\ref{fi:2}b) those that are peaked
away from 0.
All the resonances that can be produced by DPE (i.e. those with I~=~0
and G parity
positive) fall into the first class while those with I~=~1 and/or G parity
negative fall into the second class which
may be due to the fact that,
in part, they are produced by diffraction
from the proton vertex rather than from being due to real
central events.
Fig.~\ref{fi:1}b) shows the
$d\sigma/dM_{X_0}$,
distribution for $|x_F|$~$<$~0.1. As can be seen
the $\omega$, which can not be produced by DPE and which has a $x_F$
distribution peaked away from zero, is suppressed in the $|x_F|$~$<$~0.1
distribution relative to the total sample.
\par
Another variable that can be measured is the four momentum transfer squared
($|t|$) at one of the proton vertices. The $|t|$ distribution
has been measured for all the resonances.
All the distributions
fall into three distinct classes as summarised in
fig.~\ref{fi:2}c)-e).
Fig.~\ref{fi:2}c) (type I) has the classic shape associated with
$\pi$ exchange~\cite{phiangpap} and has been fitted by
\[
\frac{d\sigma}{dt}=\alpha \exp{b_1t} + \beta t \exp{b_2t}
\]
where $\alpha$ and $\beta$ are found to be similar in size.
Fig.~\ref{fi:2}d) (type II)
is similar to spin flip distributions
and can be fitted by
\[
\frac{d\sigma}{dt}= \alpha t \exp{b_2t}
\]
Fig.~\ref{fi:2}e) (type III) represents non-spin flip distributions and
can be fitted by
\[
\frac{d\sigma}{dt}=\alpha \exp{b_1t}
\]
The results of the fit to $d\sigma/dt$ and the $t$ distributions types
for all the
resonances observed are given in table~\ref{summary}.
\par
In addition to the kinematical variables discussed above, in central
production there are variables that are specific to the
central vertex at which the
two
exchanged particles interact to form the final state $X^0$.
The WA91 collaboration published a paper~\cite{wa91corr}
showing that
there was a difference in
the resonances observed when
the angle between the outgoing slow and fast protons was near to 0 degrees
compared to when the angle was near to 180 degrees.
In order to try to explain this effect
in terms of a physical model,
Close and Kirk~\cite{closeak}
proposed that the data be analysed
in terms of the parameter $dP_T$, which is the
difference in transverse momentum
between the particles exchanged from the
fast and slow vertices.
\par
The WA102 collaboration has presented studies of how different
resonances are produced as a function of the
parameter $dP_T$~\cite{ourpubs}.
The fraction of each resonance
has been calculated for
$dP_T$$\leq$0.2 GeV, 0.2$\leq$$dP_T$$\leq$0.5 GeV and $dP_T$$\geq$0.5 GeV.
In addition, the ratio of production at small $dP_T$ to large $dP_T$ has
been calculated.
The results are presented in table~\ref{frac} for all the resonances
observed in the WA102 experiment.
Figure~\ref{fi:3} shows the ratio of the number of events
for $dP_T$ $<$ 0.2 GeV to
the number of events
for $dP_T$ $>$ 0.5 GeV for each resonance considered.
It can be observed that all the undisputed $q \overline q$ states
which can be produced in DPE, namely those with positive G parity and $I=0$,
have a very small value for this ratio ($\leq 0.1$).
Some of the states with $I=1$ or G parity negative,
which can not be produced by DPE,
have a slightly higher value ($\approx 0.25$).
However, all of these states are suppressed relative to the
interesting states, which could have a gluonic component, which have
a large value for this ratio.
Fig.~\ref{fi:1}c) shows the
$d\sigma/dM_{X_0}$
distribution for $dP_T$~$<$~0.2 and
fig.~\ref{fi:1}d)
the $d\sigma/dM_{X_0}$ distribution
for $dP_T$~$>$~0.5.
The dominant feature of the
distribution for $dP_T$~$<$~0.2 is the low mass peak which comes
from the $\pi\pi$ system.
This $\pi \pi$ mass spectrum is dominated by the S-Wave
although there is a 10 \% D-wave contribution~\cite{pipipap}.
The S and D-waves both have the same $dP_T$ distributions
which are similar to those observed for the $f_0(980)$.
\par
In addition to the $dP_T$ dependencies,
an interesting effect has been observed in
the azimuthal angle $\phi$ which is defined as the angle between the $p_T$
vectors of the two outgoing protons.
Historically it has been assumed that the Pomeron, with ``vacuum quantum
numbers", transforms as a scalar and hence
that the $\phi$ distribution
would be flat for resonances
produced by DPE.
The $\phi$ dependences observed~\cite{ourpubs}
are clearly not flat and considerable variation
is observed among the resonances produced.
The $\phi$ distributions for all the resonances observed in the
WA102 experiment are shown in figs.~\ref{fi:4},~\ref{fi:5}.
These azimuthal dependences,
as a function of
$J^{PC}$,
are very striking.
For the mesons that can be produced by DPE
the $\phi$ distributions
maximise around $90^{o}$
for resonances with $J^{PC}$~=~$0^{-+}$,
at $180^o$ for those with
$J^{PC}$~=~$1^{++}$ and
at $0^o$
for those with
$J^{PC}$~=~$2^{-+}$.
As can be seen from fig.~\ref{fi:5} the $\phi$ distributions are not
simply a J-dependent effect, since
$0^{++}$ production peaks at $0^o$ for some states whereas others are more
evenly spread;
moreover, $2^{++}$ established $q\bar{q}$ states peak at
$180^o$ whereas the $f_2(1910)$ and $f_2(1950)$
peak at $0^o$.
\par
Several theoretical papers have been published on these
effects~\cite{angdist,clschul}.
All agree that the exchanged particle
must have J~$>$~0
and that J~=~1
is the simplest explanation for the observed $\phi$ distributions.
Close and Schuler~\cite{clschul} have calculated the $\phi$ dependences
for the production of resonances with different $J^{PC}$
for the case where the exchanged particle is a
Pomeron that transforms like a non-conserved
vector current.
In fact this is the only model in which detailed
calculations are made for resonances with spins other than
$J^{PC}$~=~$0^{-+}$.
In order
to gain insight into the nature of the particles
exchanged in central $pp$ interactions
Close, Kirk and Schuler~\cite{galuga} have compared
the predictions of this model with the data
for resonances with different $J^{PC}$ observed in the
WA102 experiment.
They found that for the production of $J^{PC}$~=~$0^{-+}$ mesons
they could predict the $\phi$ dependence and
the vanishing cross section as $t \rightarrow 0$ absolutely and fit the
$t$ slope in terms of one parameter,
$b_T$.
For the $J^{PC}$~=~$1^{++}$ mesons
they could predict the general form for the $\phi$
distribution. By fitting the $t$ slope they obtained the parameter
$b_L$; this then gave a parameter free prediction for
the variation of the $\phi$ distribution as a function of $t$ which
agrees with the data.
In addition, they obtained
absolute predictions for the $t$ and $\phi$ dependences of the
$J^{PC}$~=~$2^{-+}$ mesons which were again in accord with the data
when helicity 1 dominance was imposed.
In the $0^{++}$ and $2^{++}$ sector they were able to
fit all the $\phi$ distributions
with one parameter ($\mu^2$) and they found that
it is primarily the sign of this quantity that
drives the $\phi$ dependences.
Understanding the dynamical origin of this sign is now a central issue
in the quest to
distinguish $q \overline q$ states from
glueballs or other exotic states.

\par
In a recent paper by Close and Kirk~\cite{scalars}
studying the mixing of the scalar glueball with the
nearby $q \bar{q}$ states it was noted that the
$\phi$ dependencies of the $f_0(1370)$, $f_0(1500)$
and $f_0(1710)$,
and hence the sign of $(\mu^2)$,
could be explained if they were due to the relative phase
between the $n \bar{n}$ and $gg$ in their wavefunctions.
\par
One interesting feature is that there is very little evidence
for radially excited states in central $pp$ collisions.
Therefore it is even more striking that
there are three tensor states
in the region of 2~GeV which is where the
tensor glueball is predicted to be by Lattice QCD~\cite{lattQCD}.
Two of these states have similar $\phi$ dependencies
and one the opposite. This effect is very similar to what was observed in the
scalar sector.
It is interesting to speculate whether the $f_2(1910)$, $f_2(1950)$ and
$f_2(2150)$ are due to mixing between a tensor glueball and
nearby $q \bar{q}$ states. This may
explain their
observation in
central production as being through their gluonic component.
\par
In summary,
a study of the
production of resonances in
central $pp$ interactions has been presented.
Cuts on the
$dP_T$ variable select out known $q \bar{q}$ states from
glueball candidates. Why this works is still to be understood.
The azimuthal angle $\phi$ has given information on the
nature of the Pomeron. This information may in turn be used in future
to extract information on the gluonic content of the observed mesons.
It is interesting to note that although there is strong evidence
for all the ground state nonet states
(except for those with $J^{PC}$~=~$1^{+-}$) there is little
evidence for radially excited nonet members. This may give special
significance to the observation of the
$f_2(1910)$, $f_2(1950)$ and
$f_2(2150)$.

\begin{center}
{\bf Acknowledgements}
\end{center}
\par
This work is supported, in part, by grants from
the British Particle Physics and Astronomy Research Council and
the British Royal Society.
The author wishes to thank the WA76, WA91 and WA102 collaborations for the
use of their data.

\newpage
\begin{table}[h]
\caption{Summary of resonance production.
$t$ type I, II and III refer to the shape of the $|t|$ distribution as
depicted in fig.~\ref{fi:2}c), d) and e) respectively.
The error quoted represents the statistical and systematic errors
summed in quadrature.}
\label{summary}
\begin{center}
\begin{tabular}{|c|c|c|c|c|c|c|} \hline
 & & & & & & \\
$J^{PC}$ & Resonance&$\sigma$ (nb) &$t$ type
& Value of $b_1$ in & Value of $b_2$ in & DPE\\
 & & at $\sqrt{s}$=29.1& & $\exp{-b_1t}$ & $t\exp{-b_2t}$ & Compatible\\
 & & GeV & & (GeV$^{-2}$) & (GeV$^{-2}$) &\\
 & & & & & & \\ \hline
 & & & & & & \\
$0^{-+}$
% &$\pi^0$  &22 011 $\pm$ 3 267 &  $\pi$ exchange b= 3.2$\pm$0.2& \\
&$\pi^0$  &22 011 $\pm$ 3 267 &I&  17.3 $\pm$ 0.2 & 17.6 $\pm$ 0.3 & \\
& $\eta$  &3 859$\pm$ 368  &II &  & 8.2$\pm$0.3 &   \\
& $\eta^\prime$  &1 717 $\pm$ 184 &II &   & 8.1$\pm$0.4 & Yes\\
 & & & & && \\ \hline
 & & & & && \\
$0^{++}$
&$a_{0}(980)$  &638 $\pm$ 60  &III& 6.2 $\pm$ 0.8 & & \\
& $f_{0}(980)$  &5 711 $\pm$ 450 &III&  5.3 $\pm$ 0.2& & Yes\\
& $f_{0}(1370)$  & 1 753 $\pm$ 580 &III& 6.7 $\pm$ 0.2& &\\
& $f_{0}(1500)$  & 2 914$\pm$ 301 &III& 5.2 $\pm$ 0.5  &&Yes \\
& $f_{0}(1710)$  & 245$\pm$ 65 &III& 6.5 $\pm$ 0.5 && \\
& $f_{0}(2000)$  &3 139 $\pm$ 480 &III& 5.6 $\pm$ 0.4 & &\\
 & & & & && \\ \hline
 & & & & && \\
$1^{++}$
&$a_{1}(1260)$  &10 011 $\pm$ 900 &III& 6.9 $\pm$ 0.2& &\\
& $f_{1}(1285)$  &6 857 $\pm$ 1 306&III&6.3 $\pm$ 0.3  & & Yes\\
& $f_{1}(1420)$  &1 080 $\pm$ 385 &III& 5.6 $\pm$ 0.5  & & Yes\\
 & & & & && \\ \hline
 & & & & && \\
$1^{--}$
& $\rho(770)$  &3 102 $\pm$ 250 &III&5.8 $\pm$ 0.1 & & No\\
% & $\omega(782)$ &7 440 $\pm$ 553 &III& $\pi$ exchange  b= 0.65 $\pm$ 0.05 &
%%\\
& $\omega(782)$  &7 440 $\pm$ 553 &I& 3.8 $\pm$ 0.2 & 24.5 $\pm$ 0.6&\\
& $\phi(1020) $  &60$\pm$ 21 &III& 7.8 $\pm$ 1.0  && No\\
&  & & & && \\ \hline
&  & & & && \\
$2^{-+}$
% & $\pi_{2}(1670)$  &1 505 $\pm$ 145 & $\pi$ exchange b = 1.1 $\pm$ 0.2  & \\
& $\pi_{2}(1670)$  &1 505 $\pm$ 145 &I& 4.3 $\pm$ 0.2  & 22.3 $\pm$ 1.0&\\
& $\eta_{2}(1645)$  &1 907 $\pm$ 152 &II&  & 7.3 $\pm$ 1.3&\\
& $\eta_{2}(1870)$  &1 940 $\pm$ 185 &II & & 8.3 $\pm$ 2.0&\\
 & & & & && \\ \hline
 & & & & && \\
$2^{++}$
& $a_{2}(1320)$  &1 684 $\pm$ 134 &III& 7.8 $\pm$ 0.3  & &\\
& $f_{2}(1270)$  &3 275 $\pm$ 422 &II&  & 8.7 $\pm$ 0.4 & Yes \\
& $f_{2}^\prime(1520)$  &68 $\pm$ 9 &II&  & 7.7 $\pm$ 3.3 & \\
& $f_{2}(1910)$  & 528 $\pm$ 40 &III &4.9 $\pm$ 1.6  & & \\
& $f_{2}(1950)$  &2 788 $\pm$ 175 &III& 5.9 $\pm$ 0.4  & & Yes\\
& $f_{2}(2150)$  & 121 $\pm$ 12 &III &5.4 $\pm$ 1.4  & & \\
 & & & & & & \\ \hline
\end{tabular}
\end{center}
\end{table}
\begin{table}[h]
\caption{Resonance production as a function of $dP_T$
expressed as a percentage of its total contribution.
The error quoted represents the statistical and systematic errors
summed in quadrature.}
\label{frac}
\begin{center}
\begin{tabular}{|c|c|c|c|c|c|} \hline
 & & & &  & \\
$J^{PC}$ & Resonance&$dP_T$$\leq$0.2 GeV & 0.2$\leq$$dP_T$$\leq$0.5 GeV
&$dP_T$$\geq$0.5 GeV&R\\
 & & & & & \\ \hline
 & & & & & \\
$0^{-+}$
&$\pi^0$  &12 $\pm$ 2 & 45 $\pm$ 2 &43 $\pm$ 2 &0.27 $\pm$ 0.05\\
& $\eta$  &6 $\pm$ 2  & 34 $\pm$ 2 &  60 $\pm$ 3 &0.10 $\pm$ 0.03 \\
& $\eta^\prime$  &3 $\pm$ 2 & 32 $\pm$ 2  &64 $\pm$ 3 & 0.05 $\pm$ 0.03\\
 & & & & &\\ \hline
 & & & & &\\
$0^{++}$
&$a_{0}(980)$  &25 $\pm$ 4  & 33 $\pm$ 5 &42 $\pm$ 6 & 0.57 $\pm$ 0.14\\
& $f_{0}(980)$  &23 $\pm$ 2 & 51 $\pm$ 3 &26 $\pm$ 3 &0.88 $\pm$ 0.12\\
& $f_{0}(1370)$  &18 $\pm$ 4 & 32 $\pm$ 2 &50 $\pm$ 3 & 0.36 $\pm$ 0.08\\
& $f_{0}(1500)$  &24$\pm$ 2 & 54$\pm$ 3 &22 $\pm$ 4& 1.05 $\pm$ 0.18\\
& $f_{0}(1710)$  &26 $\pm$ 2 & 46 $\pm$ 2 &28  $\pm$ 2 & 0.95 $\pm$ 0.10\\
& $f_{0}(2000)$  &12 $\pm$ 2 & 38 $\pm$ 3&50$\pm$ 4 & 0.23 $\pm$ 0.04\\
 & & & & &\\ \hline
 & & & & &\\
$1^{++}$
&$a_{1}(1260)$  &13 $\pm$ 3 & 51 $\pm$ 4 &36  $\pm$ 3 &0.36 $\pm$ 0.09\\
& $f_{1}(1285)$  &3 $\pm$ 1& 35 $\pm$ 2 &61 $\pm$ 4 & 0.05 $\pm$ 0.02\\
& $f_{1}(1420)$  &2 $\pm$ 2 & 38 $\pm$ 2 &60 $\pm$ 4 & 0.03 $\pm$ 0.03\\
 & & & & &\\ \hline
 & & & & &\\
$1^{--}$
& $\rho(770)$  &6 $\pm$ 2 & 41 $\pm$ 4 &53 $\pm$ 3 & 0.12 $\pm$ 0.05\\
& $\omega(782)$  &10 $\pm$ 2 & 40 $\pm$ 2 &49 $\pm $ 3 & 0.20 $\pm$ 0.04\\
& $\phi(1020) $  &8$\pm$ 3 & 47 $\pm$ 3 &45  $\pm$ 4& 0.18 $\pm$ 0.07\\
&  & & & &\\ \hline
&  & & & &\\
$2^{-+}$
& $\pi_{2}(1670)$  &11 $\pm$ 2 & 48 $\pm$ 4  &40 $\pm$ 4& 0.27 $\pm$ 0.06 \\
& $\eta_{2}(1645)$  &9 $\pm$ 1 & 32 $\pm$ 3 &59 $\pm$ 5 &0.15 $\pm$ 0.02\\
& $\eta_{2}(1870)$  &8 $\pm$ 1  & 29 $\pm$ 3 &63 $\pm$ 6 &0.13 $\pm$0.02\\
 & & & & &\\ \hline
 & & & & &\\
$2^{++}$
& $a_{2}(1320)$  &10 $\pm$ 2 & 38 $\pm$ 5 &52 $\pm$ 6 & 0.19 $\pm$0.05\\
& $f_{2}(1270)$  &8 $\pm$ 1 & 29 $\pm$ 1 &63  $\pm$ 2 & 0.12 $\pm$ 0.02\\
& $f_{2}^\prime(1520)$  &4 $\pm$ 3 & 36 $\pm$ 3 &60  $\pm$ 4 & 0.07 $\pm$
0.04\\
& $f_{2}(1910)$  &20 $\pm$ 4 & 62 $\pm$ 7 &18 $\pm$ 4 & 1.1 $\pm$ 0.3\\
& $f_{2}(1950)$  &27 $\pm$ 2 & 46 $\pm$ 5 &27 $\pm$ 2 & 1.0 $\pm$ 0.12\\
& $f_{2}(2150)$  &3 $\pm$ 3 & 53 $\pm$ 4 &44 $\pm$ 3 & 0.06 $\pm$ 0.08\\
 & & & & &\\ \hline
\end{tabular}
\end{center}
\end{table}
% \addtocounter{figure}{-13}
\clearpage
{ \large \bf Figures \rm}
\begin{figure}[h]
\caption{
The differential cross section,
$d\sigma/dM_{X_0}$,
of the centrally produced system $X^0$ for
a) all data,
b) $|x_F|$~$<$~0.1,
c) $dP_T$~$<$~0.2 GeV and
d) $dP_T$~$>$~0.5 GeV.
}
\label{fi:1}
\end{figure}
\begin{figure}[h]
\caption{
The $|x_F|$ distribution for
a) the $f_2(1270)$  and
b) the $\omega$.
The $|t|$ distributions for
c) the $\pi^0$,
d) the $\eta^\prime$ and
e) the $f_1(1285)$.
}
\label{fi:2}
\end{figure}
\begin{figure}[h]
\caption{
The ratio of the number of events with
$dP_T$~$\leq$~0.2 to the number of events with
$dP_T$~$\geq$~0.5~GeV for all the resonances observed
grouped in $J^{PC}$.
}
\label{fi:3}
\end{figure}
\begin{figure}[h]
\caption{
The azimuthal angle $\phi$ between the fast and
slow protons for
resonances with $J^{PC}$~=~$0^{-+}$, $1^{--}$, $1^{++}$ and
$2^{-+}$.
}
\label{fi:4}
\end{figure}
\begin{figure}[h]
\caption{
The azimuthal angle $\phi$ between the fast and
slow protons for
resonances with $J^{PC}$~=~$0^{++}$ and
$2^{++}$.
}
\label{fi:5}
\end{figure}
\newpage
\begin{center}
\epsfig{figure=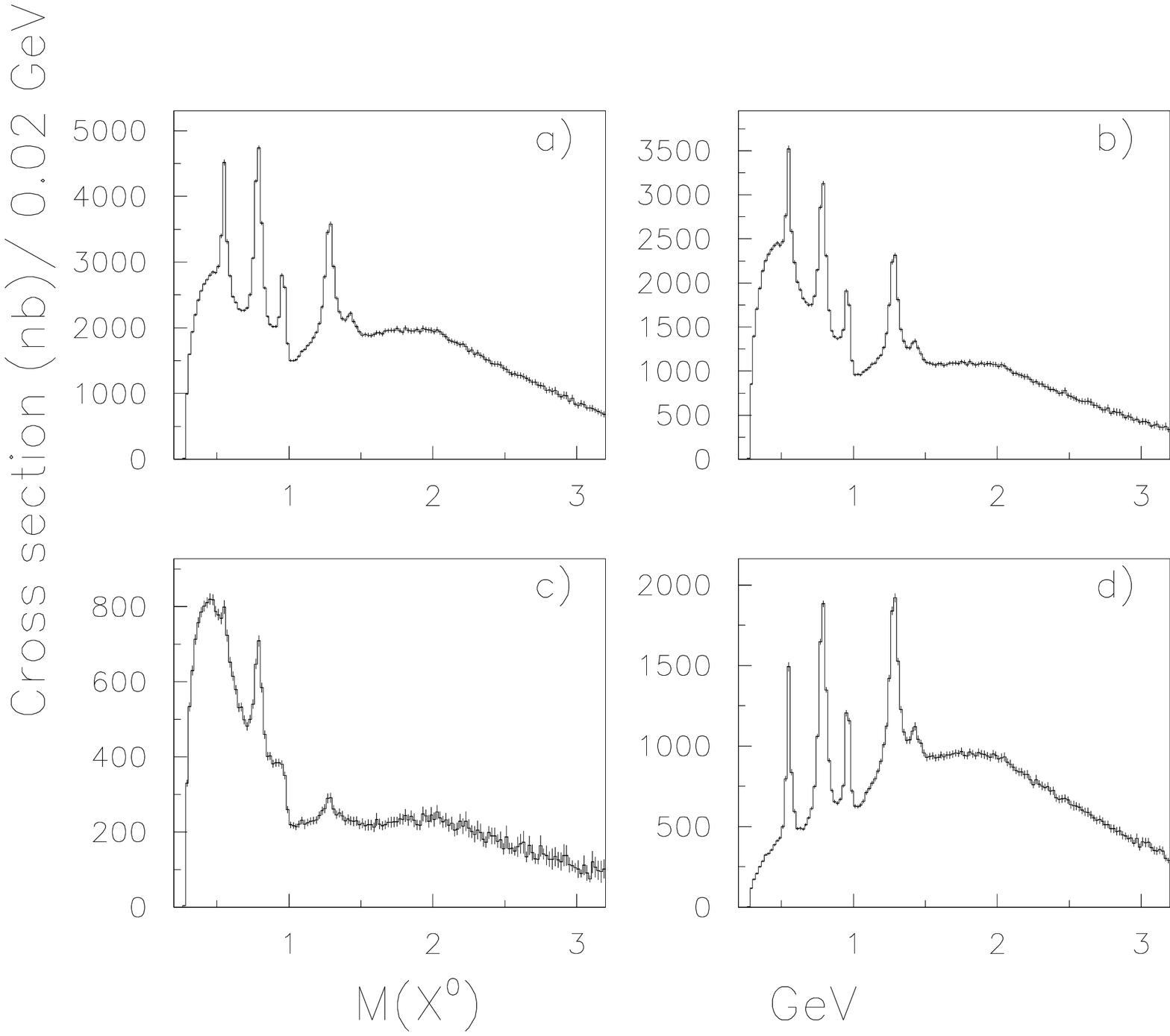,height=20cm,width=17cm}
\end{center}
\begin{center} {Figure 1} \end{center}
\newpage
\begin{center}
\epsfig{figure=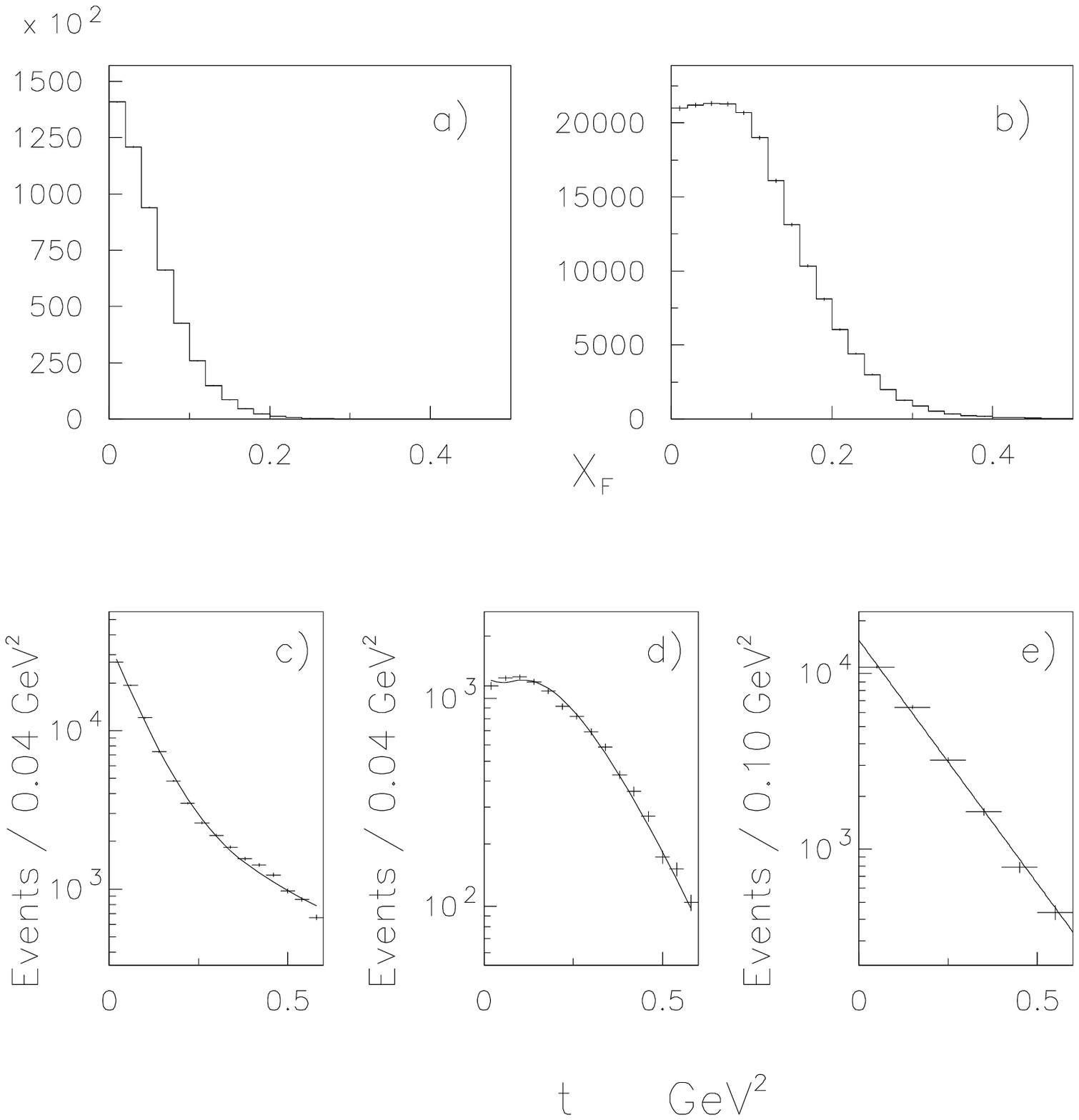,height=22cm,width=17cm}
\end{center}
\begin{center} {Figure 2} \end{center}
\newpage
\begin{center}
\epsfig{figure=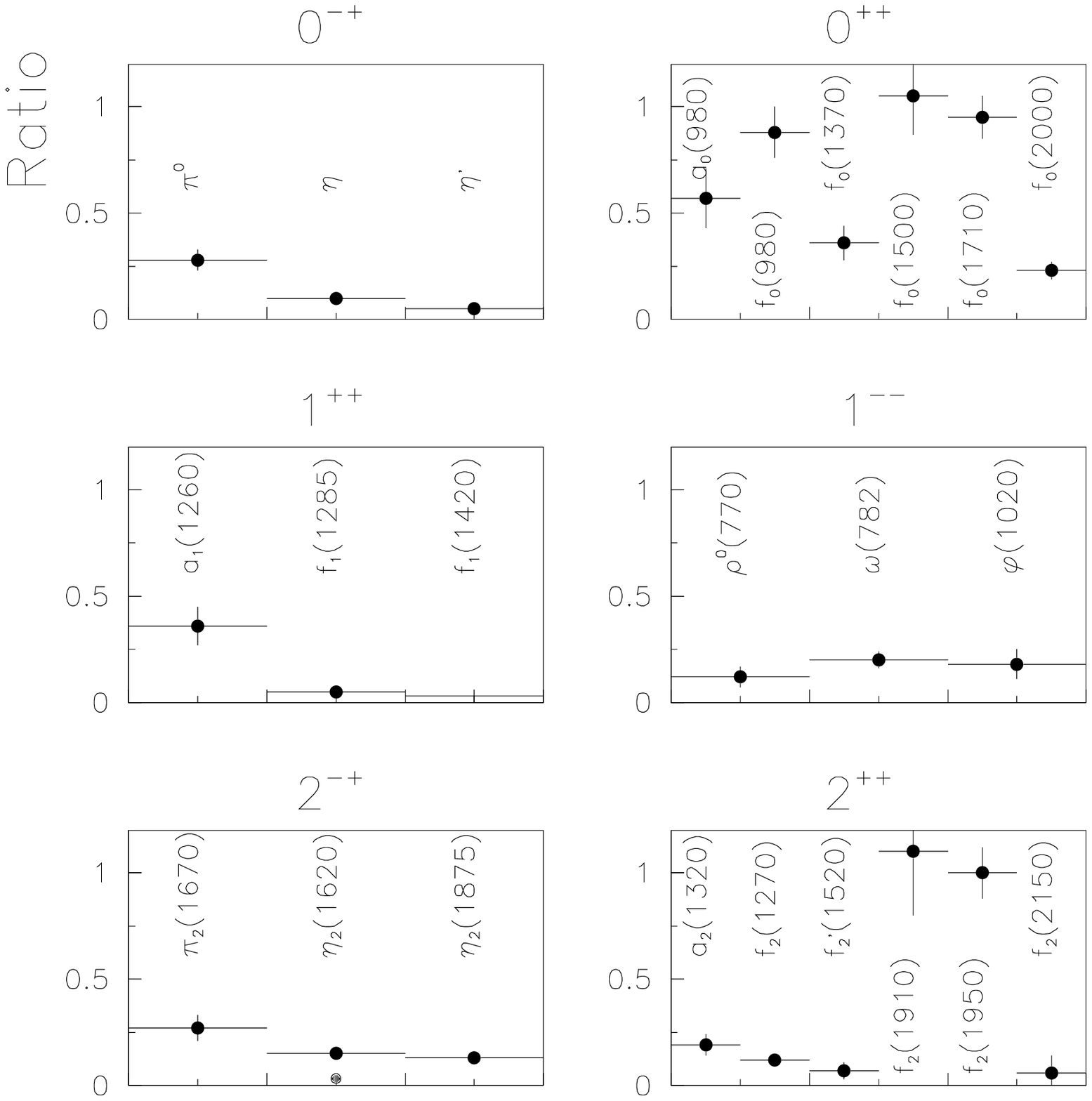,height=22cm,width=17cm}
\end{center}
\begin{center} {Figure 3} \end{center}
\newpage
\begin{center}
\epsfig{figure=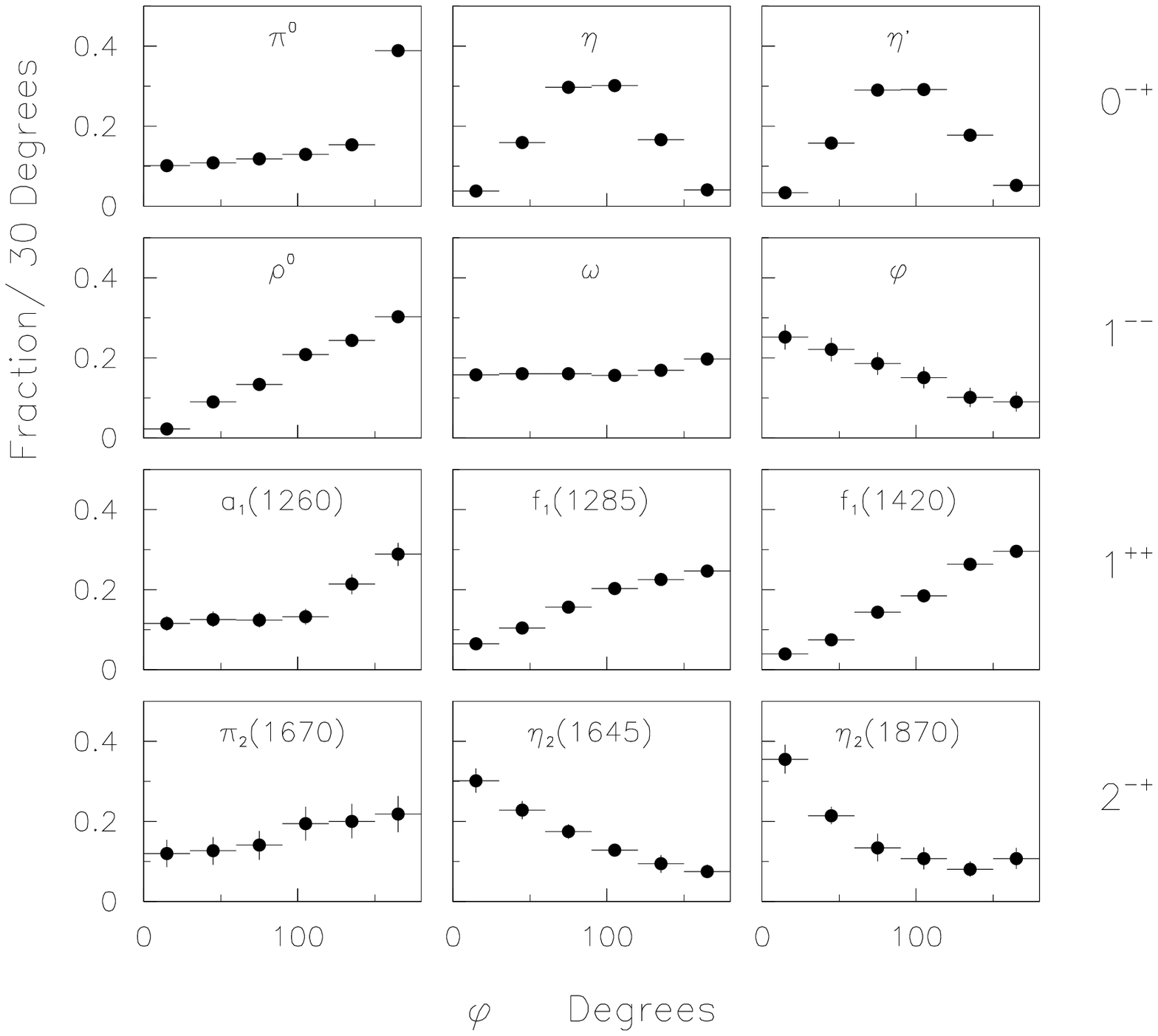,height=22cm,width=17cm}
\end{center}
\begin{center} {Figure 4} \end{center}
\newpage
\begin{center}
\epsfig{figure=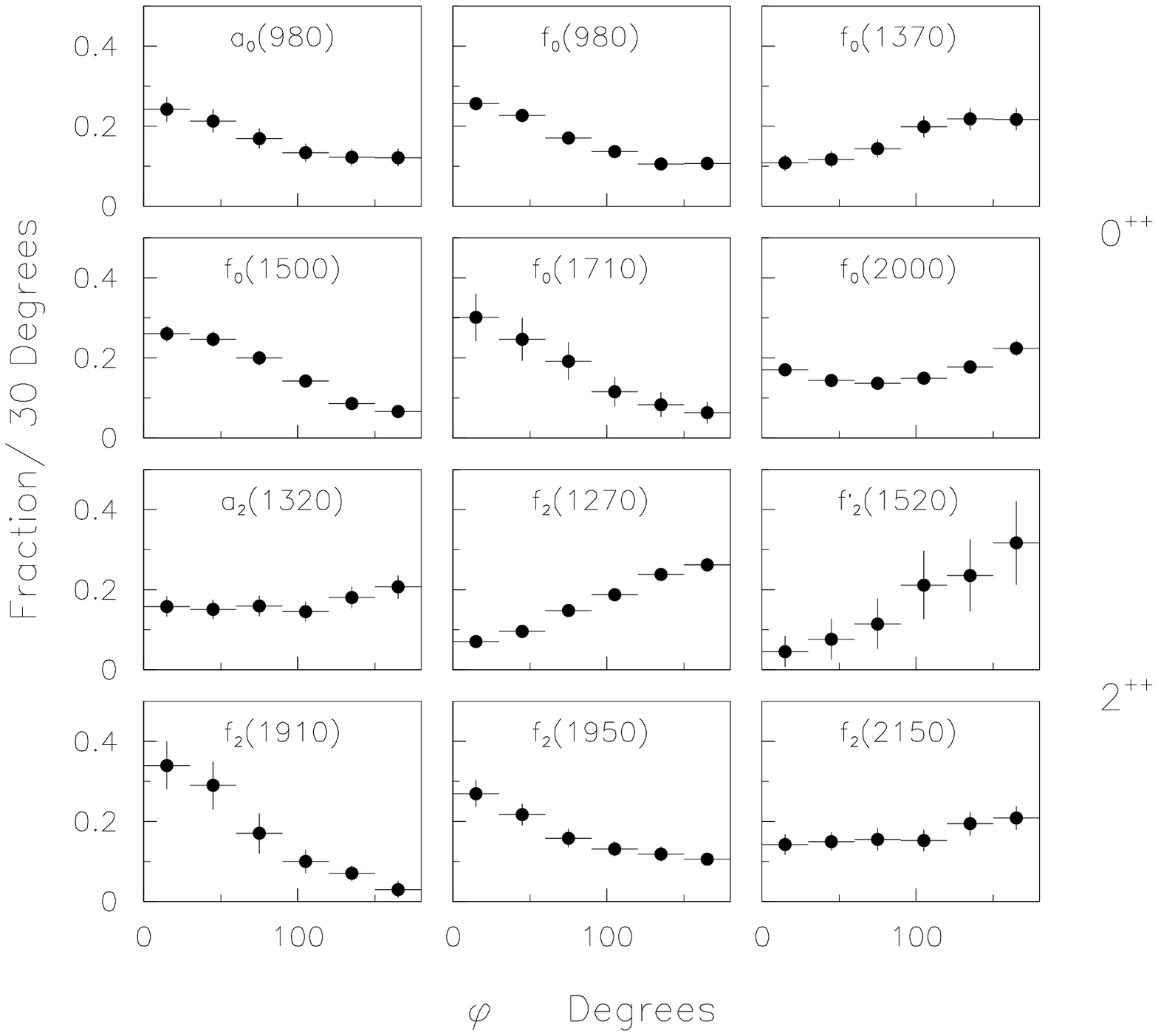,height=22cm,width=17cm}
\end{center}
\begin{center} {Figure 5} \end{center}
\end{document}